\documentclass[plb,showpacs,preprintnumbers,twocolumn,amsmath,nofootinbib]{revtex4}
\usepackage[dvips,final]{graphicx}
\usepackage{amssymb,amsmath,epsfig,bm,pifont}
\usepackage{xcolor}
\usepackage{color}
\usepackage{colordvi}
\definecolor{darkgreen}{rgb}{0,0.66,0}

\usepackage{relsize}
\usepackage[greek,english]{babel}
\usepackage[utf8]{inputenc}
\newcommand{\babar}{{\mbox{\slshape B\kern-0.1em{\smaller A}\kern-0.1em
            B\kern-0.1em{\smaller A\kern-0.2em R}}}
           }
\newcommand{\babaR}{{\mbox{\slshape B\kern-0.1em{\smaller A}\kern-0.1em
            B\kern-0.1em{\smaller A\kern-0.2em R,}}}
           }
\newcommand{\be}{\begin{equation}}\newcommand{\ee}{\end{equation}}%
\newcommand{\bd}{\begin{displaymath}}\newcommand{\ed}{\end{displaymath}}
\newcommand{\bit}{\begin{itemize}}                                        
\newcommand{\eit}{\end{itemize}}                                          
\newcommand{\ben}{\begin{enumerate}}                                      
\newcommand{\een}{\end{enumerate}}                                        
\newcommand{\baa}{\begin{array}{lll}}                                     
\newcommand{\eaa}{\end{array}}                                            
\newcommand{\ba}{\begin{eqnarray}}                                        
\newcommand{\ea}{\end{eqnarray}}                                          
\newcommand{\gev}[1]{\relax\ifmmode{\text{GeV}^{#1}}                      
                     \else{GeV$^{#1}${ }}\fi}                             
\def\MSbar{\relax\ifmmode\overline                                        
            {\rm MS}\else{$\overline{\rm MS}${ }}\fi}                     
\def\as{\relax\ifmmode \alpha_s\else{$ \alpha_s${ }}\fi}                  
\def\abar{\relax\ifmmode{\bar{a}}\else{$\bar{a}${ }}\fi}                  

\begin{document}
\thispagestyle{empty}
\date{\today}
\preprint{\hbox{RUB-TPII-03/2014}\\ }

\title{What binds quarks together at different momentum scales?
       A conceptual scenario}
\author{N.~G.~Stefanis}
\email{stefanis@tp2.ruhr-uni-bochum.de}
\affiliation{Institut f\"{u}r Theoretische Physik II,
             Ruhr-Universit\"{a}t Bochum,
             D-44780 Bochum, Germany\\}
\date{\today}

\begin{abstract}
The binding effects of quarks within hadrons are discussed in
terms of the pion distribution amplitude over longitudinal
momentum fractions.
To understand the behavior of this quantity at different momentum
scales, the concept of synchronization in complex systems has been
employed.
It is argued that at low momentum scales, the quarks get correlated
by nonlocal quark/gluon condensates that cause an endpoint-suppressed,
mainly bimodal structure of the pion distribution amplitude
inferred from a sum-rule analysis.
The mass generation mechanism, within the framework of
Dyson-Schwinger equations, and evolution effects pull these two peaks
back to the center to form at $Q^2\to\infty$ the asymptotic distribution
amplitude which represents the most synchronized $\bar{q}q$ state.
\end{abstract}

\maketitle
\noindent\textbf{1.$\;$Introduction}.
One of the greatest unsolved problems in Quantum Chromodynamics (QCD)
is the confinement phenomenon responsible for the binding effects of
partons---quarks and gluons---within hadrons
(see \cite{Brambilla:2014aaa,Kondo:2014sta} for recent reviews).
While at large momenta and energies, the color forces in the parton
interactions can be adequately and systematically described by
perturbative QCD, the regime of large distances, alias, small momenta,
cannot be treated reliably in perturbation theory.
The key for the success of perturbative QCD in the ultraviolet
domain is grounded in the fact that the strong coupling becomes
weaker as the distances between interacting partons decrease, giving
ultimately rise to an asymptotically free field
theory---``ultraviolet freedom'' \cite{Gross:1973id}.

On the other hand, the behavior of the strong coupling constant in the
infrared (IR) has not yet been formally established.
However, various calculations, based on the Dyson-Schwinger equations
(DSE), yield clear signs for the saturation of color forces in the IR,
see, e.g., \cite{Boucaud:2011ug} for a recent review.
Standard QCD perturbation theory cannot be reliably applied at low
Euclidean momenta because of the inevitable appearance of the
(unphysical) Landau singularity at momenta
$\mu^2 \sim \Lambda_\text{QCD}^2$.
Several proposals exist to rectify this problem and define an analytic
coupling in the IR---see \cite{Stefanis:2009kv} for a review.
Nevertheless, it is still unclear how the confining properties of
quarks and gluons, encoded in correlation functions, arise in
nonperturbative QCD.
Certainly, lattice calculations can provide useful benchmarks for the
confinement phase of QCD, but they have their own inherent limitations.
In search of alternatives, and without the mathematical tools to solve
QCD nonperturbatively in the continuum, we need new ideas and
organizing principles to guide us through the data in hopes of
revealing tangible predictions that can be used to test these concepts.

In this paper, I will describe a ``roadmap'' to confinement by
synthesizing different new and old ideas and methods to form a unified
perception of this phenomenon without formally solving the QCD
correlation functions in a deep mathematical sense.
Nevertheless, predictions will be presented that can be tested in
experiments in the near future.
A novelty of the approach is the use of the concept of spontaneous
synchronization of nonlinear oscillators, that has passed the test of
experiment in various areas of nonlinear science, but has never been
used before in the context of QCD.

To begin with, What are the landmarks along the confinement route?
Instead of starting at high momenta and march down to small ones, where
confinement becomes eminent for quarks, I will describe a scenario that
goes the inverse way and discuss the behavior of interlocked quarks
from low to large momentum scales.
I will expose this scenario in three steps:
(i) nonperturbative correlations,
(ii) dynamical chiral-symmetry breaking (DCSB) and mass generation, and
(iii) evolution behavior from low to (asymptotically) high momenta.
The following exposition will be basic but precise.

\noindent\textbf{2.$\;$Nonperturbative correlations}.
What binds quarks together?
Although we cannot answer this question by performing ab initio
calculations within continuum QCD, we may try to understand the salient
features of what binds quarks together at \emph{different} momentum
scales by pursuing multiple approaches and combining their results.
The primary concern in analyzing a hadronic process within QCD is how
to describe as much as possible of its dynamics in terms of hard
(i.e., short-distance) partonic subprocesses---characterized by a
large scale $Q^2$---amenable to QCD perturbation theory.
The large-distance (soft) remainder---ascribed to nonperturbative
dynamics---is then taken from experiment.
This factorization procedure becomes particularly useful, if the
isolated soft part is universal, i.e., process independent.
Using techniques from collinear factorization, a good ``laboratory''
for testing these issues is provided by the process-independent pion's
distribution amplitude (DA)
$\varphi_{\pi}(x,\mu^2)$
for finding the valence $\bar{q}q$ pair in the pion carrying the
longitudinal momentum fractions $x_q=x$ and
$x_{\bar{q}}=1-x\equiv \bar{x}$.
On the other hand, the large momentum scale $Q^2$ localizes the hard
collisions of the partons in the longitudinal direction along the
lightcone
(see \cite{Chernyak:1983ej,BL89,Stefanis:1999wy} for reviews).

The pion DA is the prototype for a two-body bound state in QCD and
is defined at the leading-twist level two by the matrix element
\begin{eqnarray}
  \langle 0| \bar{q}(z) \gamma_\mu\gamma_5 [z,0] q(0)
           | \pi(P)
  \rangle|_{z^{2}=0}
&& \!\!\!\!\! =
  if_\pi P_\mu \int_{0}^{1} dx e^{i x (z\cdot P)}
\nonumber \\
&& \times
  \varphi_{\pi}^{(2)} \left(x,\mu^2\right) \, .
\label{eq:pion-DA}
\end{eqnarray}
It is linked to the lightcone wave function of the $\bar{q}q$ pair
\cite{BL89}:
$
  \varphi_{\pi}^{(2)} \left(x,\mu^2\right)
=
  \int_{}^{\mu^{2}} \frac{d\bm{k}_{T}^{2}}{16\pi^2}
  \psi\left( x, \bm{k}_{T} \right) \, .
$
The momentum scale $\mu$ enters through the renormalization of
the current operator and denotes the maximum transverse momentum
included in the lightcone wave function of the $\bar{q}q$ pair.
We have adopted in (\ref{eq:pion-DA}) the lightcone gauge $A\cdot n=0$,
where $n^2=0$, so that the gauge link
$
 [z,0]
=
 \mathcal{P}\exp \left(
                       ig \int_{0}^{z} A^{\mu}d\tau_{\mu}
                 \right)
 =1
$.
We have also used the shorthand notation
$A^{\mu}=\sum_{a} t^{a}A^{\mu}_{a}$ ($t_{a}$ being the generators of
$SU(3)_{c}$), whereas the symbol $\mathcal{P}$ path-orders these
matrix-valued quantities along the lightlike vector $n$ from 0 to $z$.
The dependence of the pion DA on the scale $\mu$ is controlled by the
Efremov-Radyushkin-Brodsky-Lepage (ERBL) evolution equation
\cite{Efremov:1978rn,BL80}.
Though the pion DA is not directly observable, it can be used within a
factorization-based approach to calculate form factors that can be
measured in experiments.

Historically, one assumes a non-trivial vacuum that is populated by
quark
$\langle 0|\bar{q}q|0\rangle$ and gluon
$\langle 0|G^{\mu\nu}G_{\mu\nu}|0\rangle$
field condensates with a correlation length much larger than the
typical hadronic size \cite{SVZ79}.
Focusing on the quark condensate, with the fields taken at the same
point (therefore, local), this is equivalent to say that the average
virtuality is zero, corresponding to an infinite correlation length of
the vacuum fluctuation.
This concept of ``local'' vacuum condensates has been used for decades
in QCD sum rules and has provided valuable insight into the structure
of hadrons, see \cite{Chernyak:1983ej} for a review.
However, the description of dynamical quantities, such as quark
distribution amplitudes for hadrons, faces severe problems (see, e.g.,
\cite{Rad90}) that are entailed by the local character (zero-quark
virtuality) of the quark condensate.
Moreover, an infinite correlation length of the quark condensate
would lead to a cosmological constant several orders of magnitude
larger than observation \cite{Brodsky:2010xf}.

\begin{figure}
\centerline{\includegraphics[width=0.47\textwidth]{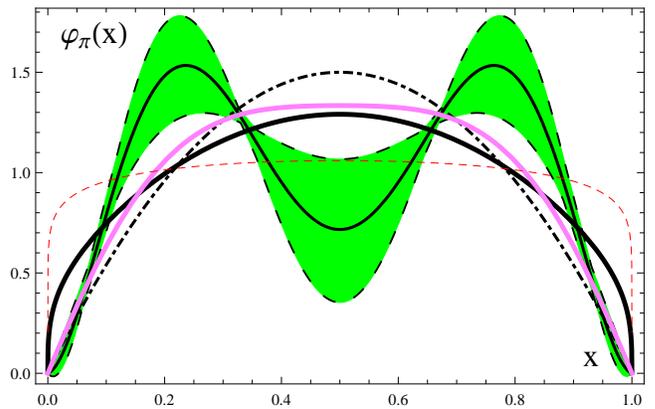}} 
\caption{(color online) Shaded (green) band contains the two-parametric BMS DAs
at $\mu^2\approx 1$~GeV$^2$ \protect\cite{BMS01}.
Curves: dashed, flat-top DA at $\mu^2=1$~GeV$^2$ from \protect\cite{MPS10};
lower solid, DSE-based approach \protect\cite{Chang:2013pq} at $\mu^2=4$~GeV$^2$;
dashed-dotted, asymptotic DA.
Upper thick (pink) solid line shows the shorttailed platykurtic DA \protect\cite{MPS14}
at $\mu^2 = 4$~GeV$^2$.
\label{fig:pi-DAs}}
\end{figure}

The use of nonlocal condensates in QCD sum rules (NLC-SR)s
was proposed by Radyushkin and collaborators quite long ago
\cite{NLC86-93,MR90,MR92}.
More recently, this approach was updated and refined by Bakulev,
Mikhailov, Stefanis (BMS) in \cite{BMS01} with the goal to extract
the twist-two pion DA at the scale $\mu^2\approx 1$~GeV$^2$ in terms
of the expansion coefficients, $a_2, a_4, a_6, a_8, a_{10}$
within the complete orthonormal basis on $x\in[0,1]$ of the Gegenbauer
polynomials
$C_{n}^{3/2}(2x-1)$ (isospin symmetry applied):
\begin{equation}
  \varphi_{\pi}^{(2)}(x,\mu^2)
=
  \varphi_{\pi}^{\rm asy}(x)
 + \sum_{n=2,4,\ldots}^{\infty} a_n (\mu^2)\psi_n(x) \, ,
\end{equation}
where
\hbox{$\varphi_{\pi}(x,\mu^2\to\infty)
=\varphi_{\pi}^{\rm asy}(x)
=6x\bar{x}$}
is the asymptotic pion DA and
\hbox{$\psi_n(x)=6x\bar{x}C_{n}^{3/2}(2x-1)$}.
Inverting the moments
\hbox{$
 \langle \xi^N \rangle
=
 \int_{0}^{1}dx (2x-1)^N \varphi_{\pi}^{(2)}(x, \mu^2)
$},
with the normalization condition
\hbox{$
 \int_{0}^{1}dx\varphi_{\pi}^{(2)}(x, \mu^2)
=
 1
$},
it was found~\cite{BMS01}
\hbox{$
 a_{2}^\text{BMS}\left(1~{\rm GeV}^2\right)
=
 (7/12)\left(5\langle \xi^2 \rangle - 1 \right)
\approx
 0.20
$},
$
 a_{4}^\text{BMS}\left(1~{\rm GeV}^2\right)
=
 (77/8)\left(\langle \xi^4 \rangle
 -(2/3)\langle \xi^2 \rangle + (1/21)\right)
\approx
 -0.14
$,
while the coefficients $a_6, a_8, a_{10}$ were also determined but
were neglected in the modeling because they were found to be
significantly smaller than the first two and bearing large
uncertainties, see \cite{BMS01,SBMP12} for details.
This DA is shown in Fig.\ \ref{fig:pi-DAs} as a solid line inside the
shaded (green) band which contains the whole family of two-parametric
pion DAs enclosed by the envelopes
$[a_{2}=0.134, a_{4}=-0.044]$
and
$[a_{2}=0.251, a_{4}=-0.207]$
(from top to bottom).
This two-parametric DA family complies with the moment values determined
from the QCD NLC-SR with nonlocal condensates at $\mu^2\approx 1$~GeV$^2$
considered in \cite{BMS01} and yields values for the inverse moment
$\langle x^{-1}\rangle_{\pi}= \int_0^1 \varphi_{\pi}(x)x^{-1}dx =3(1+a_2+a_4+a_6 + \ldots)$
which comply within errors with those determined via an independent sum rule \cite{BMS01},
$\langle x^{-1} \rangle_{\pi}^{\rm BMS}=3.35\pm 0.3$.
This implies that the sum of all coefficients $a_n$ is dominated by the
contribution of $a_2$ and $a_4$, a result which lends credibility to the
BMS DA family.
Moreover, the value $a_2 =0.19 \pm 0.06$ conforms with the recent
lattice estimates of the RBC and UKQCD Collaborations \cite{Arthur:2010xf}.

As one observes from Fig. \ref{fig:pi-DAs}, one key characteristic of
this type of DAs (shaded band in green color) is that the regions at the
kinematic endpoints $x\approx 0$ and $x\approx 1$ are strongly
suppressed --- even relative to the asymptotic DA (dashed-dotted line)
\cite{MPS10}.
This suppression is entailed by the finiteness of the average quark
virtuality
$
 \lambda_{q}^{2}
=
 \langle
        \bar{q}(0)D^2 q(0)
 \rangle
 / \langle
        \bar{q}(0)q(0)
 \rangle
\simeq
 \langle
 \bar{q}igG^{\mu\nu}\sigma_{\mu\nu}q
 \rangle / 2\langle
        \bar{q}(0)q(0)
 \rangle
\approx
 [0.35-0.5]
$~GeV$^2$, where $D_{\mu}=\partial_{\mu}-ig\Sigma_{a}A_{\mu}^{a}t^{a}$
and $G^{\mu\nu}$ is the gluon-field strength tensor.
In the following, the value
$\lambda_{q}^{2}(\mu^{2}\approx 1~\text{GeV}^2)\approx 0.4$~GeV$^2$
will be used, which was determined in \cite{BMS01} with the help of the
CLEO data \cite{CLEO98} on the pion-photon transition form factor---see
\cite{BM02} for lattice estimates and references.
The parameter $\lambda_{q}^{2}$ controls the strength of the nonlocal
condensate contribution in the QCD sum rules: the larger its value,
the stronger suppressed this contribution and the closer the shape of
the pion DA becomes to the asymptotic form.
Technically speaking, the profile of the BMS DAs results from the
interplay between the perturbative contribution and the dominant
nonperturbative term due to the scalar nonlocal condensate in the
theoretical part of the QCD sum rule.
Because the latter contribution is not singular in $x$ and has a dip
around $x=1/2$, it causes a bimodal endpoint-suppressed structure of
the DA profile.

The crucial assumption underlying the nonlocality of the condensate is
that in coordinate space the correlation length
$\Lambda \sim 1/\lambda_{q}$ for a $\bar{q}q$ pair behaves like
$
 \langle
           \bar{q}(z)[z,0]q(0)
 \rangle
\sim
 \langle
           \bar{q}(0)q(0)
 \rangle
 \exp\left(
           -\lambda_{q}^{2}|z^2|/8
     \right)
$.
Such a distribution function of Gaussian fluctuations means that the
$\bar{q}q$ correlation length induced by the condensate tends to stay
within a limited range, which is about
$\Lambda \sim 0.3$~fm
(for $\lambda_{q}^{2}=0.4$~GeV$^2$ \cite{BMS01}).
At large Euclidean $z^2$, the nonlocal quark condensate decays rapidly
to zero \cite{BM02,MPS10}, so that, from a distance, the virtuality
fluctuations are ironed out and the condensate practically exists only
inside hadrons (similarly to the in-hadron condensates proposed in
\cite{Brodsky:2010xf,Brodsky:2012ku}).
If we calculate the average transverse momentum of a valence quark in
the pion with the help of the $\bar{q}q$ pair wave function
$\psi(x, \bm{k}_{T})$,
assuming again a Gaussian distribution for the intrinsic
$\bm{k}_{T}$ momenta carried by the quarks \cite{SSK99-00}, we find
$\langle \bm{k}_{T}^{2} \rangle_\text{BMS}^{1/2}\sim 0.35~\text{GeV}$
which amounts to a distance of approximately
0.6~fm~$\approx \langle r_{\pi}^2\rangle^{1/2}$
(the pion's charge radius).
This value is about the same for the values
$m_{q}=0$ and
$m_{q}\approx 0.3$~GeV,
we used, and agrees well with the estimate in \cite{Chernyak:1983ej}.
On the other hand, gluons decouple and disperse their transverse
momentum to an infinite number of gluons via their self-interactions.
Thus, the vacuum field fluctuations are much shorter than the typical
transverse size of the valence-quark state---see \cite{SSK99-00} for
details.
These findings are in line with the appearance of DCSB on a scale
$\sim 0.3$~fm, see, e.g., \cite{Schweitzer:2014nea}.

\begin{figure}
\centerline{\includegraphics[width=0.40\textwidth]{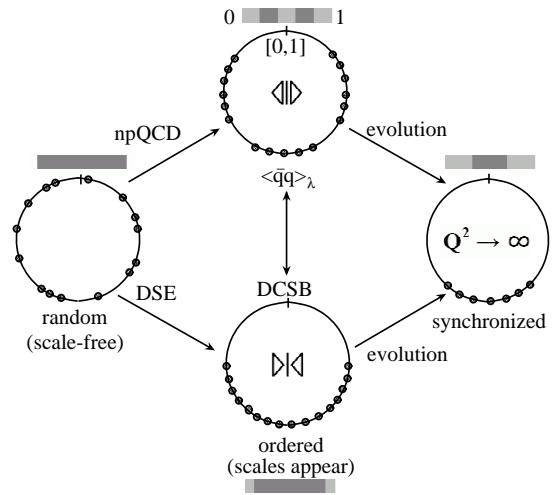}} 
\caption{(color online). Pion DAs at different momentum scales
in terms of the Kuramoto model.
The dots represent $x$ values in the interval $[0,1]$.
The tension ($\lhd\!|\!\rhd$) and compression ($\rhd\!|\!\lhd$)
tendencies in the $x$ spectrum are indicated.
The strips show the dominant $x$ regions in the corresponding DAs.
\label{fig:pi-DA-structure}}
\end{figure}

\noindent\textbf{3.$\;$Synchronization concepts}.
The most obvious characteristic of the BMS pion DA is its two-humped
structure, which is condensate-driven and reflects the tension between
the valence quark and the valence antiquark with respect to their
longitudinal momentum fractions.
To comprehend the meaning of the pion DA at different momentum scales,
it is helpful to conceive of the longitudinal momentum fractions $x$ as
being natural oscillator frequencies (phases) of a large number
($N\to\infty$)
of phase-coupled oscillators using the Kuramoto model---see
\cite{Strogatz2000} for reviews and references.\footnote{The technical
details of this model are not relevant for our qualitative exposition.}
As long as the oscillators are non-interacting, their native
frequencies---visualized in an idealized way as a swarm of points
randomly distributed on a unit circle of $x\in [0,1]$---are unlocked
building an incoherent ensemble of points
(left portrait in Fig.\ \ref{fig:pi-DA-structure}).
This situation corresponds to a constant pion DA, or a flat-top one
that vanishes at the kinematical endpoints $x=0,1$, e.g.,
$
 \varphi_{\pi}^\text{flat-top}(x)
=
 \Gamma(2(\alpha+1)) [\Gamma^{2}(\alpha+1)]^{-1} (x\bar{x})^{\alpha}
$
with $\alpha=0.1$---dashed (red) line in Fig.\ \ref{fig:pi-DAs}
\cite{MPS10}.\footnote{A ``table-like'' pion DA
$\varphi_{\pi}^\text{table}(x)\simeq (x\bar{x})^{0.05}/0.91$ with
$\lambda_{q}^{2}\sim 0.35$~GeV$^2$ was proposed in \cite{MR90}.}
Such a distribution is scale-free, meaning that no $x$ region
is singled out to be associated with a valence quark (antiquark)
because all locations on the unit circle are indistinguishable;
the pion looks like a pointlike particle without internal structure,
see, for example, \cite{Chang:2011vu}.
Besides, $\varphi_\pi(x)=\text{const}$
(corresponding to a vanishing pion charge radius)
would yield results for the electromagnetic and transition form
factors in conflict with experiment \cite{Chang:2011vu}.

The nonlocal condensate $\langle \bar{q}q\rangle_{\lambda}$, as a
clear manifestation of nonperturbative QCD,
creates a color-singlet proto-pion and
causes the set of the $x$ values in the pion DA to flock into two
distinct clusters: one close to $x\approx 0.75$, the other at
$x\approx 0.25$.
These clusters correspond to two groups of synchronized oscillators
(upper graph in Fig.\ \ref{fig:pi-DA-structure}),
whereas the endpoints $x=0,1$ around the ``North pole'' are almost
depleted.
This pattern conforms with the generic profile of a BMS-like DA in
Fig.\ \ref{fig:pi-DAs}.
It suggests that most configurations of the valence $\bar{q}q$ pair
tend to have either a leading quark or a leading antiquark, though
configurations in which the valence quark and the valence antiquark
share comparable fractions of the longitudinal momentum of the pion
around $x=1/2$ are also possible but are less favorable.
In accordance with Fig. \ref{fig:pi-DAs}, the size of the two clusters
bears large uncertainties.\footnote{Inclusion of more coefficients
$a_n$ would eventually entail more and smaller clusters.}
The same applies to the region around the ``South pole'' in
Fig.\ \ref{fig:pi-DA-structure}, which corresponds to the central
region $x=1/2$ in Fig. \ref{fig:pi-DAs}, while the absence
of dots around the ``North pole'' is quite strict.
This is, because in our approach \cite{BMS01} the uncertainties on the
shape of the $\pi$DA in the endpoint regions $x=0,1$ are very small
(see Fig. \ref{fig:pi-DAs}).

Note that the well-known Chernyak-Zhitnitsky DA \cite{Chernyak:1983ej}
would correspond to a pattern (not shown) with two distinct clusters
concentrated at the endpoints $x=0,1$, while the central region $x=1/2$
would be almost empty.

\noindent\textbf{4.$\;$DCSB and mass generation}.
The other important feature of confinement is DCSB and the generation
of quark and gluon masses.
At a deeper level of understanding of the QCD dynamics in the IR,
it is likely that condensate formation and mass generation are intertwined
phenomena.
However, at present it is prudent to discuss these effects separately
using specific schemes.
An appropriate framework to study the mass-generation effects is
provided by the DSE-based method, see \cite{Cloet:2013jya} for a recent
review.
The dressed-quark mass $\sim 0.3$~GeV converts the real quark pole
in the dressed quark propagator into a complex one, whereas the
\emph{effective} gluon mass, with a dressed-gluon mass scale in the
range 0.4-0.6~GeV \cite{Cloet:2013jya}, enters the argument of the
strong coupling and provides saturation of the color forces in the IR.
In Fig.\ \ref{fig:pi-DAs} we show a pion DA---solid (blue)
line---obtained with the DSE methodology
\cite{Chang:2013pq,Cloet:2013jya,Maris:1997tm,Gao:2014bca}---Eq. (15)
in \cite{Chang:2013pq}.
It derives from the nonperturbative content of the Bethe-Salpeter
kernels in the dressed quark and gluon propagators associated with
DCSB, the later being exclusively responsible for the broadening of
this DA relative to $\varphi_{\pi}^\text{asy}$ \cite{Chang:2013pq}.
At the renormalization point $\mu=2$~GeV it is described by the concave
function
$
 \varphi_{\pi}^{\rm DSE}(x)=1.81(x\bar{x})^{a}
 [1+\tilde{a}_{2}C_{2}^{a+1/2}(2x-1)]
$
with
$a=0.31$, $\tilde{a}_{2}=-0.12$.
A similarly downward concave DA,
$
 \varphi_{\pi}^{\rm AdS/QCD}(x)=(8/\pi)(x\bar{x})^{1/2}
$,
was derived within a holographic approach to QCD embedded in a
five-dimensional Anti-de Sitter (AdS) space \cite{AdS}.
Using the Kuramoto model, the portrait of $\varphi_{\pi}^{\rm DSE}$
is represented by the graph at the bottom in Fig.\
\ref{fig:pi-DA-structure}.
It describes a pack of partially synchronized oscillators with
natural frequencies in a wide range of $x$ values.
Similar considerations apply to the AdS/QCD DA.

\begin{figure}[t]
\centerline{\includegraphics[width=0.47\textwidth]{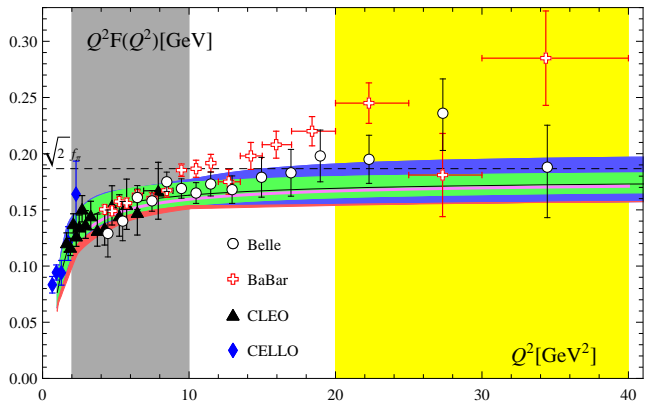}} 
\caption{(color online) Scaled pion-photon TFF vs. $Q^2$ in
comparison with data.
The designations are given in the text.}
\label{fig:scaled-TFF}
\end{figure}

Thus, the ``true'' pion DA seems to be determined by the balance of
two competing effects: the correlation caused by the
$\langle \bar{q}q\rangle_{\lambda}$
condensate, pushing the $x$ values away from the center at $x=1/2$ to
form a two-cluster arrangement, and DCSB which tends to enhance
the central $x$ region by broadening the shape of the DA and
create---speaking in terms of the Kuramoto-model
analogy---a single moderately synchronized group of oscillators
(see Fig.\ \ref{fig:pi-DA-structure}).
One might argue that at low scales, $\mu \approx 2$~GeV,
$
 \varphi_{\pi}^\text{true}(x)
\approx
 a\varphi_{\pi}^\text{BMS}(x) + (1-a)\varphi_{\pi}^\text{DSE}(x)
$.\footnote{This would imply that at nonperturbative scales
$\psi^\text{true}$ is a superposition of $\psi^\text{BMS}$ and
$\psi^\text{DSE}$ (or AdS/QCD).}
This synthesized DA would have features pertaining to both confinement
facets, exhibiting profile characteristics inherited from both DAs:
endpoint suppression like $\varphi_{\pi}^{\rm BMS}$ and central-region
enhancement like $\varphi_{\pi}^{\rm DSE}$.
For $a\approx 0.7-0.9$, it would still belong to the family of BMS-like
DAs shown in Fig.\ \ref{fig:pi-DAs} within the shaded area
yielding an inverse moment
[$\langle x^{-1}\rangle_{\pi}
=3(1+a_{2}+a_{4} \ldots)
=
3/(\sqrt{2}f_{\pi})Q^2F_{\gamma^*\gamma\pi^0}^{(\rm LO)}(Q^2)
$]
with values in the range
$
 \langle x^{-1}\rangle_{\pi}^{\rm true}
\sim
 \langle x^{-1}\rangle_{\pi}^{\rm BMS}
\lesssim 3.5 < \langle x^{-1}\rangle_{\pi}^{\rm DSE} \approx 4.6$
and, as a result, a pion-photon transition form factor (TFF) inside
the  margin of predictions in Fig.\ \ref{fig:scaled-TFF}.
The accurate determination of the mixing parameter $a$, which controls
the tradeoff between the endpoint suppression and the broadness of
the $\pi$DA, will be discussed separately in a future publication.
Here suffice it to say that one may select within the BMS scheme a DA
which is a downward concave curve over a broad interval
of $x$ values but which still exhibits endpoint suppression
entailed by the nonlocal condensate \cite{MPS14}.
This short-tailed platykurtic pion DA belongs to a family of
admissible DAs derived with the nonlocality $\lambda_{q}^{2}=0.45$~GeV$^2$
and is displayed in Fig.\ \ref{fig:pi-DAs} (thick solid pink line).
The close resemblance between this DA and the DSE one is obvious.
But the distinct behavior from the DSE DA at the endpoints is key in
deriving predictions for the pion-photon TFF in good agreement with the
data (see Fig. \ref{fig:scaled-TFF}).
As the pion DA evolves to higher $Q^2$, QCD interactions die out and
the DA reaches at $Q^2\!\to\!\infty$ its asymptotic form which
represents the most synchronized $\mathbf{\bar{q}q}$ configuration
(Fig.\ \ref{fig:pi-DA-structure}).

\noindent\textbf{5.$\;$Litmus test of the approach}.
The scenario exposed above, can be tested experimentally by measuring
the pion-photon TFF
$F^{\gamma^{*}\gamma\pi^{0}}(q_{1}^{2}=Q^2,q_{2}^{2}\to 0)$
with
$Q^2 \gg \Lambda_\text{QCD}^{2}$.
This is the gold-plated QCD observable because it arises from the
factorization properties of QCD, with all binding nonperturbative
effects being absorbed into the twist-two and twist-four pion DAs.
Hence, the $Q^2$ behavior of this TFF reflects and reveals the
underlying structure of the pion DA.
The calculation of $F^{\gamma^{*}\gamma\pi^{0}}$ has been carried
out within our approach---based on lightcone sum rules (LCSR)s
\cite{BBK89,Kho99,SY99}---in \cite{BMS02} and subsequently in
\cite{BMS05-12}, with technical details being provided in \cite{SBMP12}.
The TFF within the method of LCSRs is calculated with the help of
Eq.\ (2) in \cite{SBMP12} using the expressions provided in App. A
and App. B in the same reference.
The upshot of this calculation are state-of-the-art predictions,
shown in Fig.\ \ref{fig:scaled-TFF}.
The broad horizontal (green) band represents the TFF which uses as
input the two-parametric family of BMS DAs, discussed above, and includes
the NLO perturbative corrections, i.e., $T_{\rm LO}$ and $T_{\rm NLO}$,
as well as the twist-four term in terms of an effective twist-four DA
\cite{Kho99}, while the
main next-to-next-to-leading order contribution, $T_{\rm NNLO}$, is
taken into account together with the twist-six term \cite{ABOP10}
in the form of uncertainties (see \cite{SBMP12} for details).
ERBL evolution is also included at NLO.
The narrower (blue) strips above and below the broader (green) one
show the influence of the uncertainties induced by the next higher
coefficient $a_6$, while the very narrow (red) strip, at lower $Q^2$
values, represents the effect on the calculated TFF of a non-vanishing
virtuality of the quasireal photon caused by the untagged electron
in the Belle experiment \cite{Belle12} with the value
$q_{2}^{2}\approx 0.04$~GeV$^2$, as detailed in \cite{SBMP12}.
The pink line just below the BMS one (central line of the band)
denotes the prediction obtained with the platykurtic DA and has similar
statistical accuracy with respect to the data.
The presented predictions for the endpoint-suppressed DAs of
our approach agree very well with all existing data that are compatible with
QCD scaling:
CELLO \cite{CELLO91}, CLEO \cite{CLEO98}, and Belle \cite{Belle12}.
The same applies to the BaBar data \cite{BaBar09} below 9~GeV$^2$.
However, there is no matching between our scaling predictions and
the auxetic behavior of the high $Q^2$ BaBar data above 10~GeV$^2$
\cite{SBMP12}.

%
There are two momentum regimes for the pion-photon TFF which will be
probed experimentally by two different collaborations in the near
future---see Fig.\ \ref{fig:scaled-TFF}.
Window I (shaded area towards the left):
Measurement data with high statistics in the spacelike
region $2< Q^2 < 10$~GeV$^2$, taken with the BESIII
(Beijing Spectrometer) detector at the BEPC-II
(Beijing Electron Positron Collider) facility, in
$e^+e^\rightarrow \pi^+\pi^-J/\Psi$
collisions can be used to study TFFs of light mesons \cite{U12BESIII}.
Window II (shaded area towards the right):
Single-tagged measurements of the pion TFF will be performed with the
Belle II detector at the upgraded KEKB accelerator (SuperKEKB) in
Japan in the next few years and are expected to cover a wide range of
momenta up to about 50~GeV$^2$, where the data is much sparser.
A confirmation of the predictions in Fig.\ \ref{fig:scaled-TFF} will
provide a key piece of evidence for the presented
approach.

\noindent\textbf{6.$\;$Conclusions}.
In conclusion, I have aggregated different concepts and methods
together in order to provide insight into the inner structure of the
$\bar{q}q$ component of the pion DA as it appears at different momentum
scales from the typical hadronic domain to the asymptotic regime.
While the binding effects at low momenta are mainly due to nonlocal
condensates, combined with mass dressing owing to DCSB, at very high
momentum the quarks in the pion are in lockstep only as a result of
synchronization.

\noindent\textbf{Acknowledgments}.
I would like to thank Sergey Mikhailov and Alexander Pimikov for
collaboration and discussions.


\end{document}